# SWEET: Serving the Web by Exploiting Email Tunnels

Amir Houmansadr    Wenxuan Zhou    Matthew Caesar    Nikita Borisov
University of Illinois at Urbana-Champaign
Urbana,IL
{ahouman2, wzhou10, caesar, nikita}@illinois.edu

## Abstract

Open communication over the Internet poses a serious threat to countries with repressive regimes, leading them to develop and deploy censorship mechanisms within their networks. Unfortunately, existing censorship circumvention systems do not provide high *availability* guarantees to their users, as censors can identify, hence disrupt, the traffic belonging to these systems using today's advanced censorship technologies. In this paper we propose SWEET, a highly available censorship-resistant infrastructure. SWEET works by encapsulating a censored user's traffic to a proxy server inside email messages that are carried over by public email service providers, like Gmail and Yahoo Mail. As the operation of SWEET is not bound to specific email providers we argue that a censor will need to block all email communications in order to disrupt SWEET, which is infeasible as email constitutes an important part of today's Internet. Through experiments with a prototype of our system we find that SWEET's performance is sufficient for web traffic. In particular, regular websites are downloaded within couple of seconds.

## 1. INTRODUCTION

Today's Internet provides users with an environment to freely communicate, and to exchange ideas and information with others from around the world. However, free communication continues to threaten repressive regimes, as the open circulation of information and speech among their citizens can pose serious threats to their existence. Recent unrest in the middle east demonstrates that the Internet can be widely used by citizens under these regimes as a very powerful tool to spread censored news and information, inspire dissent, and organize events and protests. As a result, repressive regimes extensively monitor their citizens' access to the Internet and restrict open access to public networks [38] by using different technologies, ranging from simple IP address blocking (e.g., through access control lists) and DNS hijacking to more complicated and resource-intensive Deep Packet Inspection (DPI) [4, 25].

With the use of censorship technologies, a number of different systems were developed to retain the openness of the Internet for the users living under repressive regimes [3, 8, 13, 17, 22]. These systems are composed of an ensemble of network hosts and use different computer and networking technologies to evade the monitoring and blocking performed by the censors. The earliest circumvention tools are HTTP proxies [1, 8, 13] that simply intercept and manipulate a client's HTTP requests, defeating IP address blocking and DNS hijacking techniques. The use of more advanced censorship technologies such as deep packet inspection [4, 14], rendered the use of HTTP proxies ineffective for circumvention. This led to the advent of more advanced circumvention tools such as Ultrasurf [3] and Psiphon [22], designed to evade content filtering performed by the more advanced censors. In addition to special-purpose anti-censorship systems, many users have been using anonymity systems as effective tools to evade Internet censorship [4, 25]. These systems are designed to hide a user's activity over the Internet, which can also help to evade Internet censorship since the censor will not be able to determine the network destination of a user's traffic. Multiple designs have been proposed for anonymity systems, including the onion routing [35] and mix networks [11].

While these circumvention tools have helped, they face several challenges that prevent them from being a good choice for a longer-term solution to Internet censorship. We believe that the biggest challenge to existing circumvention systems is their lack of *availability*, meaning that a censor can disrupt their service frequently or even disable them completely [19, 27, 29, 30, 33]. The common reason leading to the mentioned lack of availability is that the network traffic made by these systems can be distinguished from regular Internet traffic by censors, i.e., such systems are not *unobservable*. This enables censors to disrupt/block the communications made by their citizens to such circumvention systems. For example, the popular Tor [17] network works by having users connect to an ensemble of nodes with public IP addresses, which proxy the user's traffic to the requested, censored destinations. This public knowledge about Tor's IP addresses, which is required to make Tor usable by users globally, can be/is used by censors to block their citizens from accessing Tor [5, 36]. To improve their availability, recent proposals for circumvention aim to make their traffic unobservable from censors [9, 16, 18, 20, 24, 37]. Several designs [9, 16, 18] seek unobservability by sharing secret information with their clients, which are not known to censors. For instance, the Tor network has recently adopted the use of *Tor Bridges*, a set of volunteer nodes connecting clients to the Tor network, whose IP addresses are selectively distributed among Tor users by Tor. Unfortunately, this approach poses another challenge [27, 30], which is sharing such secret information only with real users in a scalable manner such that it is not disclosed to the censors pretending to be users. A more recent approach in designing unobservable, hence highly-available, circumvention systems is to integrate censorship circumvention with the Internet infrastructure [20, 21, 24, 37]. Telex [37], Cirripede [20], and LAP [21] are example designs that suggest modifications to Internet infrastructure, e.g., routing decisions, in order to hide users' circumvented traffic from their monitoring censors. Even though such systems provide better availability promises, compared to traditional circumvention, their deployment requires substantial modifications to ISP networks, requiring cooperation from ISP operators and/or network equipment vendors, presenting a substantial deployment challenge.

In this paper, we design and implement SWEET, a censorship circumvention system that provides high availability by leveraging the openness of email communications. Figure 1 shows the main architecture of SWEET. A SWEET client, confined by a censoring ISP, tunnels its network traffic with blocked destinations inside a series of email messages that are exchanged between the client and an email server operated by SWEET's server. The SWEET server, then, acts as an Internet proxy [26] for the client by proxying the encapsulated traffic to blocked Internet destinations. The SWEET client uses an oblivious, public mail provider (e.g., Gmail, Hotmail, etc.) to exchange the encapsulating emails, rendering standard email filtering mechanisms ineffective in identifying/blocking SWEET-related emails. More specifically, to use SWEET for circumvention a client creates an email account with *any* public email provider (e.g., Gmail, Hotmail) and obtains SWEET's client software from an out-of-bound channel (similar to other circumvention systems). The user configures the installed SWEET software to use her public email account, which sends/receives encapsulating emails messages on behalf of the user to/from the publicly known email address of SWEET, e.g., `tunnel@sweet.org`. Note that there is no need for the user to obtain any secret information, secret key, or secret design information in order to use SWEET.

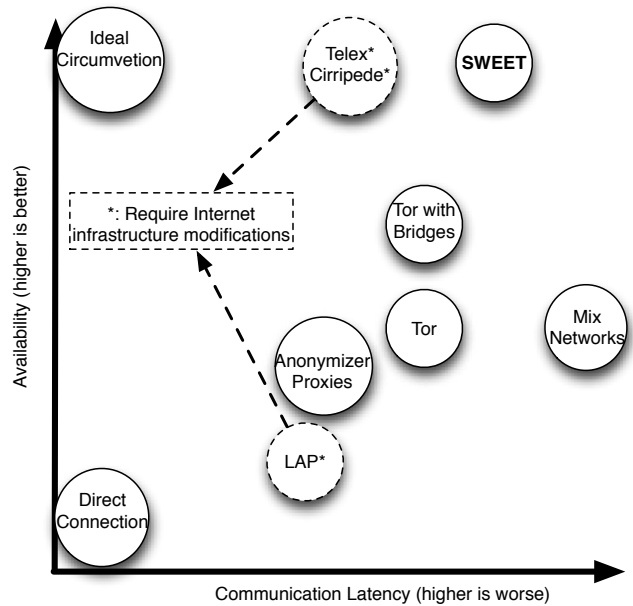

Figure 2: Comparing availability and communication latency of several circumvention systems.

SWEET provides several key advantages as compared to the existing circumvention systems. First, since email is an essential service in today's Internet it is very unlikely that a censorship authority will block *all* email communications to the outside world, due to different financial and political reasons. This, along the fact that SWEET can be reached through *any* email service, provides a high degree of *availability* for SWEET since a censor will need to block all email traffic to the Internet in order to block SWEET. Second, by using encrypted email messages SWEET is highly unobservable from the censors. Third, the real-world deployment of SWEET does not require cooperation of any third-party entity, e.g., an ISP, a web destination, or even an specific email provider. Finally, unlike several recent proposals [9, 16, 18, 24] a SWEET user does not have to obtain any secret information in order to use SWEET, providing high user convenience and ensuring the security and privacy of the user.

In fact, the high availability of SWEET comes for the price of higher, but bearable, communication latencies. Figure 2 compares SWEET with several popular circumvention systems regarding their availability and communication latency. As our measurements in Section 7 show, SWEET provides communication latencies that are convenient for latency-sensitive activities like web browsing (i.e., few seconds). Such additional, tolerable latency of SWEET comes with the bonus of better availability, as discussed in Section 5.2.

We have built a prototype implementation for the SWEET system and evaluated its performance. We have also prototyped two different designs for SWEET's client software, as proposed in this paper. The first client design uses email protocols, e.g., POP3 and SMTP, to communicate with the SWEET system, and our second design is based on using the webmail interface. Our measurements show that a SWEET client is able to browse regular-sized web destinations with download times in the order of couple of seconds.

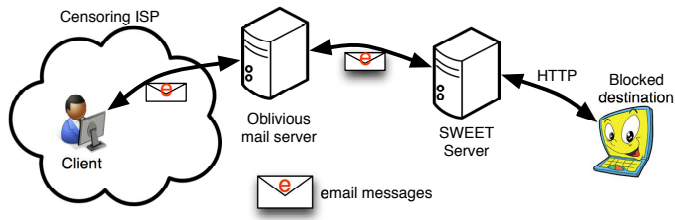

Figure 1: Overall architecture of SWEET.

In summary, this paper makes the following main contributions: i) we propose a novel infrastructure for censorship circumvention, SWEET, which provides high availability, a feature missing in practical circumvention systems; ii) we develop two prototype implementations for SWEET (one using webmail and the other using email exchange protocols) that allow the use of nearly all email providers by SWEET clients; and, iii) we show the feasibility of SWEET for practical censorship circumvention by measuring the communication latency of SWEET for web browsing using our prototype implementation.

The rest of this paper is organized as follows; in Section 2, we provide some background information and discuss the related work on censorship circumvention. In Section 3, we reviews our threat model. We provide the detailed description of the proposed circumvention system, SWEET, in Section 4. We discuss SWEET's censorship features, including its availability, in Section 5 and compare it with the literature. Our prototype implementation and evaluations are presented in Sections 6 and 7, respectively. Finally, we conclude the paper in Section 8.

## 2. RELATED WORK

As a result of extensive censorship of the Internet by repressive regimes, affected citizens have been looking for effective tools to gain unrestricted access to the Internet. Early censors used simple blocking techniques such as IP address blocking and DNS hijacking; hence, the early circumvention tools are based on proxying the traffic to the blocked destinations, i.e., by using an HTTP proxy [15]. Examples of proxy-based circumvention tools include Anonymizer [8], Freenet [13] and Ultrasurf [3], that helped a number of users to bypass the Internet censorship in the early days of Internet censorship. Proxying network traffic is also adopted by the Tor anonymous communication network [17] to help users bypass the censorship. Tor bridges [16] proxy the censored clients' traffic to the Tor network. The main challenge with the Tor bridges and other proxy-based circumvention systems is that keeping the IP address of the proxies unknown to the censors is a challenging problem [19, 27, 29, 30, 33]; a censor learning the IP address of the proxies can easily block any access to them and also identify their users.

The use of more advanced technologies by the censors, e.g., content filtering using deep packet inspection, resulted in the emergence of more complicated circumvention systems [9, 18, 20, 24, 37]. These systems aim in protecting their availability by hiding the use of their systems from the censors in the first place. As an example, Infranet [18] shares a secret key and some secret URL addresses with a client, which is then used to establish an unobservable communication between the client and the Infranet system, thereby enabling access the blocked destination. As another example, Collage [9] works by having a client and the Collage system *secretly* agree on some user-generated content sharing websites, e.g., flickr.com, and communicate using steganography. Unfortunately, sharing secret information with a wide range of clients is a serious challenge for these systems, as a censor can obtain the same secret information by pretending to be a client.

Some recent research suggests circumvention being built into the Internet infrastructure to better provide unobservability [20, 24, 37]. These systems rely on collaboration from some Internet routers that intercept users' traffic to uncensored destinations to establish covert communication between the users and the censored destinations. This provides a high degree of unobservability: a client's covert communication with a censored destination appears to the censor to be benign traffic with a non-prohibited destination. Telex [37] and Cirripede [20] provide this unobservable communication without the need for some pre-shared secret information with the client, as the secret keys are also covertly communicated inside the network traffic. Cirripede [20] uses an additional client registration stage that provides some advantages and limitations as compared to Telex [37] and Decoy routing [24] systems. Even though these systems are a large step forward in providing unobservable censorship circumvention their real-world deployment is highly dependent on collaboration from a number of ISPs/ASes, bringing into question whether they will be deployed in the near future.

**SWEET-like systems.** There are two projects that work in a similar manner to SWEET: the foe-project [2] and the MailMyWeb [28] for-profit service. Instead of tunneling traffic, which is the case in SWEET, these systems simply download a requested website and send it as an email attachment to the requesting user. This highly limits therir performance compared to SWEET, as discussed in Section 4.4.

## 3. THREAT MODEL

We assume that a user is confined inside a censoring ISP, e.g., a user living under a repressive regime. The censoring ISP blocks the user's access to certain Internet destinations, namely *blocked destinations*. We assume that the censor uses today's advanced filtering technologies, including IP address blocking, DNS hijacking, and deep packet inspection techniques [25]. The censoring ISP also monitors all of its egress/ingress traffic to detect any use of circumvention techniques by users that try to evade the censorship.

We assume that the censoring ISP's censorship is constrained not to degrade the *usability* of the Internet. In other words, even though the censoring ISP *selectively* blocks certain Internet connections, she is not willing to block key Internet services *entirely*. In particular, the operation of SWEET system relies on the fact that a censoring ISP does not block *all* email communications, even though she can selectively block email messages/email providers. We also assume that the censoring ISP has as much information about SWEET as any SWEET client (SWEET does not share any secret information with its clients).

We also consider an active behavior for the censoring ISP. An active censor, in addition to traffic monitoring, manipulates its egress/ingress Internet traffic, e.g., by selectively dropping some packets, and adding additional latency to some packets, in an attempt to disrupt the use of circumvention systems and/or to detect the users of such systems. Again, such perturbations are constrained to preserve the usability of the Internet for benign users.

## 4. DESIGN OF SWEET

In this section, we describe the detailed design of SWEET, our email-based censorship circumvention system. Figure 1 shows the overall architecture of SWEET. SWEET tunnels network connections between a client and a server, called SWEET server, inside email communications. The assumption that a censor does not block *all* email communications, as stated in our threat model in Section 3, ensures a strong availability for SWEET since it does not rely on any specific email provider. Upon receiving the tunneled network packets, the SWEET server acts as a transparent proxy between the client and the network destinations requested by the client. In the following, we describe the detailed description of SWEET's client and server architectures.

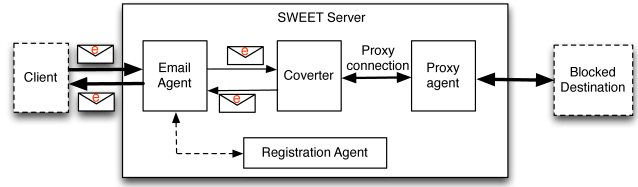

Figure 3: The main architecture of SWEET server.

### 4.1 SWEET server

The SWEET server is a computer server running outside the censored region. The SWEET server helps a censored user to evade censorship by proxying her traffic to blocked Internet destinations. More specifically, the SWEET server communicates with a censored user by exchanging email messages that carry network packets of the user's tunneled traffic. Figure 3 shows the building blocks of the SWEET server, which is composed of the following four main components:

① **Email agent:** The email agent component is an IMAP and SMTP server that receives email messages that contain the tunneled Internet traffic, being sent by SWEET clients to SWEET's publicized email address `tunnel@sweet.org`. The email agent passes the received email messages to another components of the SWEET server, i.e., the converter agent and the registration agent, to get processed accordingly. The email agent also sends email messages to SWEET clients, which are generated by other components of SWEET server and contain tunneled network packets or client registration information.

② **Converter:** The converter component processes the emails passed by the email agent, and extracts the tunneled network packets. The converter, then, forwards the extracted data to another component of the SWEET server, the proxy agent component. Also, the converter component receives network packets from the proxy agent and converts them into email messages that are targeted to the email address of corresponding SWEET clients. The converter component then passes these email messages to the email agent for delivery to their intended recipients. As described later, the converter encrypts/decrypts the email attachments of a user using a secret key shared with that user.

③ **Proxy agent:** The proxy agent proxies the network packets of SWEET clients that are extracted by the converter component, and sends them to the Internet des-tination requested by the clients. In other words, the proxy agent makes a proxy connection with SWEET clients, being tunneled inside the email-based communication. Through the established proxy connections, the client requests access to Internet destinations, e.g., blocked web sites.

④ **Registration agent:** This component is in charge of registering the email addresses of the SWEET clients, prior to their use of SWEET. The information about the registered clients can be used to ensure quality of service for all users and prevent denial-of-service attacks on the SWEET server. Additionally, the registration agent shares a secret key with the client, which is used to encrypt the tunneled information between the client and the SWEET server.

The email agent of the SWEET server receives two type of email messages; *traffic emails*, which contain tunneled traffic from the clients (sent to tunnel@sweet.org address), and *registration emails*, which carry client registration information (sent to register@sweet.org).

**Client registration:** Before the very first use of the SWEET service, a client needs to register her email address with the SWEET system. This is automatically performed by the client's SWEET software, through the same email channel used for traffic tunneling. The objective of client registration is twofold: to prevent denial-of-service (DoS) attacks and to share a secret key between a client and the SWEET server. A denial-of-service attack might be launched on the SWEET server to disrupt its availability, e.g., through sending many malformed emails on behalf of non-existing email addresses (this is discussed in Section 5). In order to register (or update) the email address of a client, the client's SWEET software sends a registration email from the user's email address, e.g., user@gmail.com, to the SWEET's registration email address, i.e., register@sweet.org, requesting registration. The email agent forwards all received registration emails to the registration agent (④) of the SWEET server. For any new registration request, the registration agent generates and sends an email to the requesting email address (through the email agent) that contains a unique computational *challenge* (e.g., [23]). After solving the challenge, the client software sends a second email to register@sweet.org that contains the solution to the challenge, along with a Diffie-Hellman [34] public key $K_C = g^{k_C}$. If the client's response is verified by the registration agent the client's email address will be added to a *registration list*, that contains the list of registered email addresses with their expiration time. Also, the registration agent uses its own Diffie-Hellman public key, $K_R = g^{k_R}$, to evaluate a shared key $k_{C,R} = g^{k_R k_C}$ for the later communications with the client. The registration agent adds this key to the client's entry in the registration list, to be used for communications with that client. The client is able to generate the same $k_{C,R}$ key using SWEET's publicly advertised public key and her own private key [34].

**Tunneling the traffic:** Any traffic email received by the email agent is processed as follows: the email agent (①) of SWEET server forwards the received traffic email to the converter agent (②). The converter agent processes the traffic email and extracts the tunneled information from the email. The converter agent, then, decrypts the extracted traffic information (using the key $k_{C,R}$ corresponding to the user) and sends it to the proxy agent (③) of SWEET server. Finally, the proxy processes the received packet as required, e.g., sends the packet to the requested destination. Similarly, for any tunneled packet received from the proxied destinations, the proxy agent sends it to the converter agent. The converter agent encrypts the received packet(s) (using the corresponding $k_{C,R}$), and generate a traffic email that contain the encrypted data as email attachment. Each email is targeted to the email address of the corresponding client (e.g., by specifying the To: field of the email message). The generated email is passed to the email agent, who sends the email to the corresponding client. Note that to improve the latency performance of the connection, small packets that arrive at the same time get attached to the same email message.

## 4.2 SWEET client

To use SWEET, a SWEET client needs to obtain a copy of SWEET's client software and install it on her machine. The client further needs to have an email account with a public email provider, e.g., Gmail[1] mail service. The choice of an encrypted versus plaintext email service makes a tradeoff between the usage unobservability and the performance. The use of an encrypted email service, e.g., Gmail, improves the usage unobservability while using a plaintext email service, e.g., Hotmail, improves the connection's throughput; this is discussed in Section 5.1. A client needs to configure the installed SWEET's software with information about her email account. Prior to the first use of SWEET by a client, the client software registers the email address of its user with the SWEET server and obtains a shared secret key $k_{C,R}$, as described in Section 4.1.

We propose two designs for SWEET client software: a protocol-based design, which uses standard email protocols to exchange email with client's email provider, and a webmail-based design, which uses the webmail interface of the client's email provider. We describe these two designs in the following.

### 4.2.1 Protocol-based design

Figure 4 shows the main elements of the protocol-based design of SWEET client. This is composed of

---

[1]https://www.gmail.com

three main components:

❶ **Web Browser:** A SWEET client needs to use a web browser to render the web sites accessed through SWEET. The client can use any web browser that supports proxying of connections, e.g., Google Chrome, Internet Explorer, or Mozilla Firefox. The client needs to configure her web browser to use a local proxy server, e.g., by setting `localhost:4444` as the HTTP/SOCKS proxy in the browser's settings. The client can use two different web browsers for browsing with and without SWEET in order to avoid the need for frequent re-configurations of the browser. Alternatively, some browsers (e.g., Google Chrome, and Mozilla Firefox) allow a user to have multiple browsing *profile*s, hence, a user can setup two profiles for browsing with and without SWEET.

❷ **Email Agent:** This component sends and receives SWEET emails thorough the client's public email account. The client needs to configure the email agent with the settings of the SMTP and IMAP/POP3 servers of her public email account. The client also needs to provide the email client with the required login information of her email account.

❸ **Converter:** This component sits in the middle of the web browser and the email agent, and converts SWEET emails into network packets and vice versa. The converter uses the keys shared with SWEET, $k_{C,R}$, to encrypt/decrypt email contents.

Once the client enters a URL into the configured web browser (component ❶), the browser makes a proxy connection to the local port that the converter (❸) is listening on (as specified in the proxy settings of the browser). The converter accepts the proxy connection from the browser and keeps the state of the established TCP/IP connections. For packets that are received from the web browser the converter generates traffic emails, targeted to `tunnel@sweet.org`, having the received packets as encrypted email attachments (using the key $k_{C,R}$). Such emails are passed to the email agent (❷) that sends the emails to the SWEET server through the public email provider of the client (as configured).

The email client is also configured to receive emails from the client's email account through an email retrieval protocol, e.g., IMAP or POP3. This allows the email agent to continuously look for new emails from the SWEET server. Once new emails are received the email agent passes them to the converter, who in turn extracts the packet information from the emails, decrypts them, and sends them to the web browser over the existing TCP/IP connection with the browser.

*4.2.2 Webmail-based design*

As an alternative approach to the protocol-based design described above, the SWEET client can use the webmail interface of the client's public email provider to exchange emails with the SWEET server.

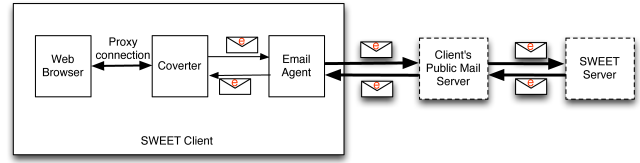

Figure 4: **The protocol-based design for SWEET client.**

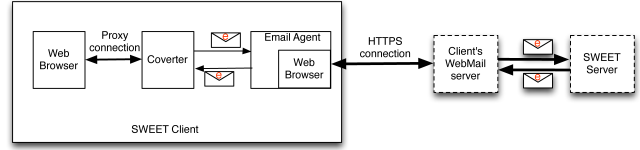

Figure 5: **The webmail-based design for SWEET client.**

Figure 5 shows the main architecture of our webmail-based design. The main difference with the protocol-based design is that in this case the email agent (component ❷) uses a web browser to exchange emails. More specifically, the email agent uses its web browser to open a webmail interface with the client's email account, using the user's authentication credentials for logging in. Through this HTTP/HTTPS connection, the email agent communicates with the SWEET server by sending and receiving emails. The rest of the webmail-based design is similar to the protocol-based design. If desired, the email agent can use the same web browser that the user uses for normal web browsing.

### 4.3 The choice of the proxy protocol

As mentioned before, the SWEET server uses a proxy agent that receives the tunneled traffic of clients and establishes connections to the requested destinations. We consider the use of both SOCKS [26] and HTTP [32] proxies in the design of SWEET, as each provides unique advantages. To provide the users with both options, our SWEET server's proxy agent runs a SOCKS proxy and an HTTP proxy in parallel, each running on a different port. A user can choose to use the type of proxy by configuring her SWEET client to connect to the corresponding port number.

The use of the SOCKS proxy allows the client to make *any* IP connection through the SWEET system, including dynamic web communications, such as Javascript or AJAX, and instant messaging services. In contrast, an HTTP proxy only allows access to HTTP destinations. However, an HTTP proxy may speed up connections to such destinations by using HTTP-layer optimizations such as caching or pre-fetching of web objects.

### 4.4 An alternative approach: Web download

A trivial approach in providing censorship circumvention using email is to download an entire webpage and attach it as an email attachment to email messages that are targeted to the requesting users. In fact, this approach is under development by the open-source foe project [2], and the for-profit service of MailMyWeb [28]. Unfortunately, this simple approach only provides a limited access to the Internet: a user can only access static websites. In particular, this approach cannot be used to access destinations that require end-to-end encryption, contain dynamic web applications like HTML5 and Javascript sockets, or need user login information. Also, this approach does not support accessing web destinations that require a live Internet connection, e.g., video streaming websites, instant messaging, etc. In fact, the MailMyWeb service uses some heuristics to tackle some of these shortcomings partially, which are privacy-invasive and inefficient. For example, in order to access login-based websites MailMyWeb requests a user to send her login credentials to MailMyWeb by email. Also, a user can request for videos hosted *only* on the YouTube video sharing website, which are then downloaded by MailMyWeb and sent as email attachments; this causes a large delay between the time a video is requested until it is has received by the user. SWEET, on the other hand, provides a comprehensive web browsing experience to its users since it can tunnel any kind of IP traffic.

## 5. DISCUSSIONS AND COMPARISONS

In this section we evaluate SWEET's circumvention capabilities by discussing important features that are essential for an effective circumvention.

### 5.1 Unobservability and the use of encryption

We say a circumvention tool provides unobservability if censors are not able to identify the users of that tool by monitoring the Internet traffic of their citizens. Unobservability has been considered in designing several recent circumvention systems [20, 24, 37] as a mechanism to ensure the availability of the service.

Usage unobservability can be a very desirable property to users living under repressive regimes, as it can reduce a user's risk of suspicion by the government. However, if desired, SWEET users may disable unobservability while retaining availability. This provides the benefit of reduced latency. It comes at the expense of observability, but users in some environments may wish to retain reliable anonymous communications yet do not fear reprisal from their government. SWEET users have the option to use SWEET in a highly unobservable manner, or to trade off unobservability for better communication performance. More specifically, the use of an encryption-enabled email service (e.g., Gmail, Hushmail[2]) with SWEET provides usage unobservability for the user, as the censor will not be able to see the recipient's email address, e.g., `tunnel@sweet.org`. On the other side, the use of plaintext email services (e.g., Yahoo!, Hotmail) provides better availability as repressive regimes occasionally disrupt or block the access to encryption-enabled services, including encrypted emails (yet they do not block all email communications). As a recent instance, Iran blocked all HTTPS connections of its citizen during the second week of February 2012 [7], the anniversary of Iran's green movement in 2009. In addition, while using an encryption-enabled email, to ensure unobservability the user's email traffic patterns should mimic that of normal email communications, e.g., to defeat traffic analysis by a censor; this limits the bandwidth available to the user, as discussed in Section 7.2.

### 5.2 Availability

SWEET's availability is tied to the assumption discussed previously that a censor is not willing to block *all* email communications, as it would degrade the usability of the Internet for its users. As the use of SWEET does not require an email account with any specific email provider, users can always find an email service to get connected to SWEET.

IP filtering and DNS hijacking techniques would not be able to stop SWEET traffic as a SWEET user's traffic is destined to her public email provider, but not to an IP address or nameserver belonging to SWEET system. Another technique used by today's sophisticated censors is deep packet inspection (DPI). The use of encryption-enabled email renders DPI ineffective, as the email headers get encrypted and the DPI will not be able to analyze the email headers in order to detect the email addresses of SWEET, to hence block the traffic. In the case of plaintext emails, to defeat DPI SWEET server can provide different *email aliases* to different users or to change its public email address frequently. Note that generating email aliases has no cost for SWEET server and can be done with no limit. In the worst case, a user can obtain her specialized email address through an out of band channel, or by connecting through a encryption-enabled email account (as mentioned before the DPI is ineffective on encryption-enabled emails).

As another approach for disrupting the operation of SWEET, a censor might try to launch a denial-of-service (DoS) attack on SWEET server. The common techniques for DoS attacks, e.g., ICMP flooding and SYN flooding, can be mitigated by protecting the SWEET server using up-to-date firewalls. Alternatively, a malicious entity (e.g., a censor) can try to exhaust the SWEET service by sending disruptive traffic through the email communication channel of SWEET. In other words, a censor can play the role of a SWEET client and send Internet traffic through its SWEET client software in a way that breaks or overloads the SWEET server. As an example, the attacker can flood the SWEET's SOCKS proxy by initiating many incomplete SOCKS connections, or sending SYN floods. A censor could send such attacking requests on behalf of a number of rogue (non-existing) email addresses, to render an email blacklist maintained by SWEET server ineffective in preventing such attacks. As a result, SWEET server should deploy effective mechanisms to protect against possible DoS attacks. One effective mechanism is to require a new user to register her email address with SWEET server prior to her first use of SWEET. Such registration can be performed in an unobservable manner by SWEET's client software through the email communication channel (see Section 4.1). Also, to ensure the quality of service for all users, the SWEET server can limit the use of SWEET by putting a cap on the volume of traffic communicated by each registered email address.

---

[2] `www.hushmail.com`

### 5.3 Other properties of SWEET

**Confidentiality:** As mentioned before, SWEET encrypts the tunneled traffic, i.e., the email attachments are encrypted using a key shared between a user and SWEET server. This ensures the confidentiality of SWEET user communications from any entity wiretapping the traffic, including the censorship authorities and the public email provider. Note that the email attachments are encrypted even if the user to choose a plaintext email service. To make a connection confidential from SWEET server the user can use an end-to-end encryption with the final destination, e.g., by using HTTPS. Alternatively, a user can use SWEET to connect to another circumvention system, like Anonymizer [10], to ensure confidentiality from SWEET server.

**Ease of deployment:** We argue that SWEET can be easily deployed on the Internet and provide service to a wide range of users. First of all, SWEET is low-cost and needs few resources for deployment. It can be deployed using a single server that runs a few light-weight processes, including a mail server and a SOCKS proxy. To service in a large scale SWEET server can be deployed in a distributed manner by several machines in different geographic locations. Secondly, the operation of SWEET is standalone and does not rely on collaboration from other entities, e.g., end-hosts or ISPs. This provides a significant advantage to recent research that relies on collaboration from ISPs [20, 24, 37], or end-hosts [9, 18]. In fact, the easy setup and low-resources of SWEET's deployment allows it to be implemented by individuals with different levels of technical expertise. For instance, an ordinary home user can deploy a personal SWEET server to help her friends in censored regions evade censorship, or a corporate network can setup such system for its agents residing in a censored country.

**User convenience:** As mentioned before, a recent study [10] surveying the use of circumvention tools in censored countries shows that users give the most preference to the ease of use when choosing a circumvention tool. The use of SWEET is simple and requires few resources from a client. A SWEET client only needs to install the provided client software, that can be obtained from out-of-band channels like social networks or downloaded from the Internet. Due to its simple design, an expert user can also develop the client software herself. In addition to SWEET software, the user needs to have an email account with a public email provider, and needs to know the public information related to SWEET, e.g., the email addresses of SWEET.

**No need to share secrets:** To ensure availability and unobservability, several circumvention tools need to share some secret information with their users, in order to initiate the circumvented a connection [9, 16, 18, 24]. This is a significant limitation, as keeping such information secret from the censorship authorities is a hard problem, and the disclosure of such secret information breaks their unobservability and availability promises. A SWEET user does not require to obtain any secret information from the SWEET server.

## 6. PROTOTYPE IMPLEMENTATION

We have prototyped the SWEET circumvention system and measured its performance.

### 6.1 Server implementation

We implement the SWEET server on a Linux machine. The machine runs *Ubuntu 10.04 LTS* and has a 2 GHz quad-core CPU and 4 GB of memory. We run Postfix[3], a simple email server that supports basic functions. Postfix listens for new emails targeted to the `register@sweet.org` and `tunnel@sweet.org` email addresses. Postfix stores the received emails into designated file directories that are continuously watched by the converter and registration agent components of SWEET server. Each stored email has a unique name, concatenating the email id of its corresponding client and an increasing counter. The converter agent is a simple Python-based program that runs in the background and continuously checks the folder for new emails. The converter also converts proxied packets, passed by SWEET's proxy, into email messages and sends them to their intended clients. For the proxy agent, we use Squid[4] as our HTTP proxy and Suttree[5] as our SOCKS proxy. Squid runs on a lo-cal port and listens for connections from the converter component.

---
[3] http://www.postfix.org/
[4] http://www.squid-cache.org/
[5] http://suttree.com/code/proxy/

## 6.2 Client implementation

We implement both protocol-based and webmail-based versions of the SWEET client.

**Protocol-based design** The client prototype is built on a desktop machine, running *Linux Ubuntu 10.04 TLS*. We set up a web browser to use the local port "localhost:9034" as the SOCKS/HTTP proxy. The converter component of SWEET client is a simple python script that listens on port 9034 for connections, e.g., from our web browser. Finally, we implement the email agent of SWEET client using *Fetchmail*[6], a popular client software for sending and retrieval of emails through email protocols. We generate a free Gmail account and configure Fetchmail to receive emails through IMAP[7] and POP3[8] servers of Gmail, and to send emails through the SMTP server of Gmail[9]. Note that our design does not rely on Gmail, and the client software can be set up with any email account.

**Webmail-based design** Our webmail-based implementation also runs on *Linux Ubuntu 10.04 TLS*. Our webmail-based implementation uses a Google Chrome browser for making connections through SWEET, configured to use the local port of "localhost:9034" as a proxy. Also, the webmail-based design uses the same converter component as the one used in the protocol-based prototype.

We prototype the web-based email agent by running a *UserScript*[10] inside the Mozilla Firefox[11] web browser. More specifically, we install a Firefox extension, called *Greasemonkey*[12], that allows a user to run her own JavaScript, i.e., Userscript, while browsing certain destinations. We write a UserScript that runs in Gmail's webmail interface and listens for the receipt of new email messages. Our UserScript saves new emails in a local directory, which is watched by the converter component. Note that the Firefox browser is directly connected to the Internet and does not use any proxies (user needs to use the configured Chrome browser to surf the web through SWEET).

## 7. EVALUATION

We evaluate SWEET using our prototype implementation, described in Section 6.

### 7.1 Performance

---
[6] http://www.fetchmail.info/
[7] https://mail.google.com/support/bin/answer.py?answer=78799
[8] https://mail.google.com/support/bin/answer.py?answer=13287
[9] https://mail.google.com/support/bin/answer.py?answer=78799
[10] http://userscripts.org/
[11] http://www.mozilla.org/en-US/firefox/new/
[12] https://addons.mozilla.org/en-US/firefox/addon/greasemonkey/

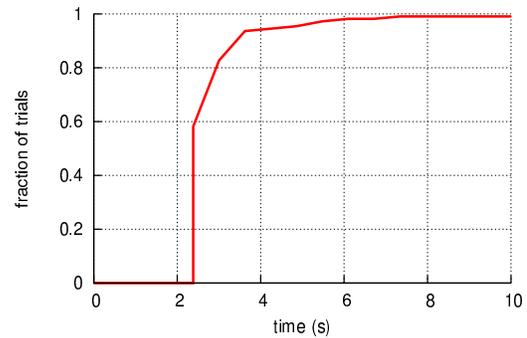

Figure 6: The CDF of the time that a SWEET email takes to travel from the SWEET client to the SWEET server (100 runs).

We use Gmail as the oblivious mail provider in our experiments. Our SWEET server is located in Urbana, IL, resulting in approximately 2000 miles of geographic distance between the SWEET server and Gmail's email server (we locate Gmail's location from its IP address). Figure 6 shows the CDF of the time that a SWEET email (carrying the tunnelled traffic) sent by a SWEET client takes to reach our SWEET server (the reverse path takes a similar time). As the figure shows, around 90% of SWEET emails take less than 3 seconds to reach the SWEET server, which is very promising considering the high data capacity of these email messages. Note that based on our measurements, most of this delay comes from email handling (e.g, spam checks, making SMTP connections, etc.) performed by the oblivious mail provider (Gmail in our experiments), but not from the network latency (the network latency and client latency constitute only tens of milliseconds of the total latency). As a result, the latency would be very similar for users with an even longer geographical distance from the oblivious mail server.

**Client registration** Before being able to request data from Internet destinations, a user needs to be registered by the SWEET server. Figure 7 shows the time taken to exchange registration messages between a client and the SWEET server. Note that the client registration needs to be performed *only once* for a long period of time. The figure shows that more than 90% of registrations establish in less than 8 seconds (with an average of 6.4 seconds).

We use two metrics to evaluate the latency performance of SWEET in browsing websites: the *time to the first appearance (TFA)* and the *total browsing time (TBT)*. As described by its name, the TFA is the time taken to receive the first response from a requested web destination. The TFA is an important metric in measuring user convenience during web browsing. For instance, suppose that a client requests a URL, e.g., http://www.cnn.com/some_news.html. By the TFA time the client receives the first HTTP RESPONSE(s) from the destination, which include the URL's text parts (perhaps the news article) along with the URLs of other objects on that page, e.g., images, ads hosted by other websites, etc. At this time the client can start reading the received portion of the website (e.g., the news article), while her browser sends requests for other objects on that webpage. On the other hand, the total browsing time (TBT) is the time after which the browser finishes fetching all of the objects in the requested URL.

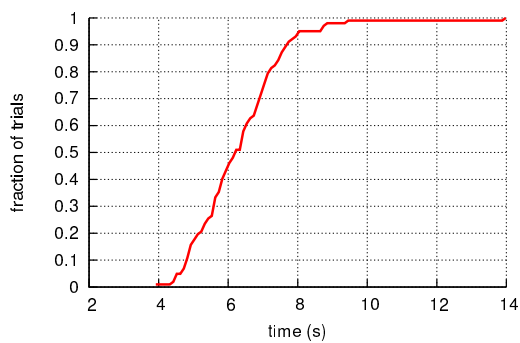

Figure 7: The CDF distribution of the registration time.

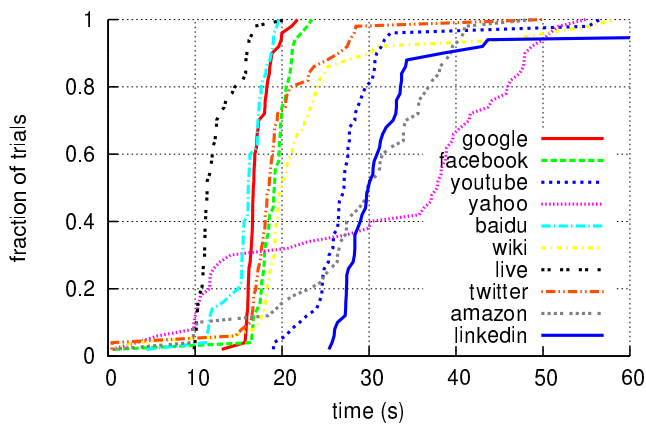

Figure 9: The CDF of the TBT time using SWEET.

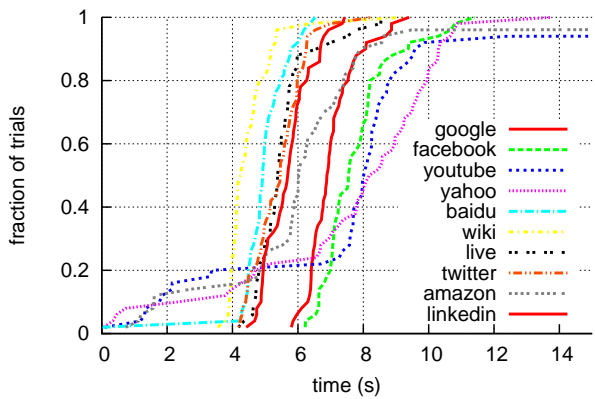

Figure 8: The CDF of the time to the first appearance (TFA) of SWEET.

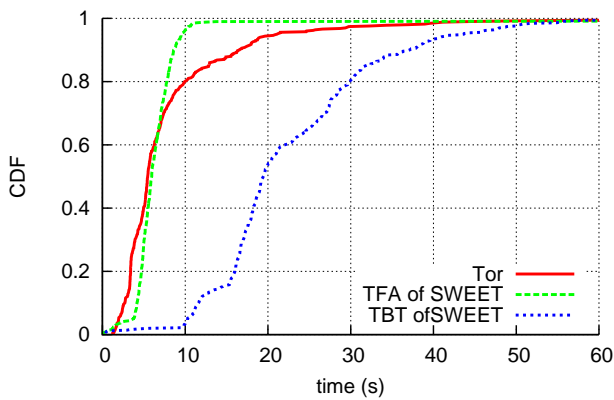

Figure 10: Comparing the average latency of SWEET and Tor.

Using our prototype we measure the end-to-end web browsing latency for the client to reach different web destinations. Figure 8 shows the time to the first appearance (TFA) using SWEET for the top 10 web URLs from Alexa's most-visited sites ranking [6]. As can be seen, the median of the TFA is about 5 seconds across all experiments, which is very promising to user convenience.

On the other hand, Figure 9 shows the total browsing time (TBT) using SWEET for the same set of destinations (50 runs for each website). As can be seen, the destinations that contain more web objects (e.g., yahoo and linkedin) take more time to get completely fetched (note that after the TFA time the user can start reading the webpage until all of the objects are received). We also run similar experiments through the popular Tor [35] anonymous network to compare its latency performance with SWEET. Figure 10 compares the cumulative time CDF for SWEET and Tor systems. As expected, our simple implementation of SWEET takes more time than Tor to browse a web page, however, it provides a sufficient performance for normal web browsing. This is in particular significant considering the strong *availability* of SWEET compared to other circumvention systems, as discussed in this paper. Additionally, we believe that further optimizations on SWEET server's proxy (like those implemented by Tor exit nodes) will further improve the latency of SWEET. Our techniques are also amenable to standard methods to improve web latency, such as plugin-based caching and compression, which can make web browsing tolerable in high delay environments [12].

## 7.2 Traffic analysis

A powerful censor can perform traffic analysis to detect the use of SWEET, e.g., by comparing a user's email communications with that of a typical email user. As a result, a SWEET user who is concerned about unobservability needs to ensure that her SWEET email communications mimic that of a normal user (as discussed in Section 5.1, a user who does not fear reprisal from her government might opt to ignore unobservability in order to gain a higher communication bandwidth). It should be mentioned that such traffic analysis is expensive for censors considering the large volume of email communications; it is estimated[13] that 294 billion email messages were sent per day in 2011.

Figure 11 shows the number of emails sent and received by a SWEET client to browse different websites. We observe that for any particular website the number of emails does not change at different runs. As can be seen, most of the web sites finish in less than three SWEET emails in each direction. The exception is the Yahoo web page as it contains many web objects, each hosted by different URLs (note that the number of email messages affects the latency performance only *sub-linearly*, since some emails are sent and received simultaneously.). Also, the average number in each way of a connection is about 4 emails. A recent study [31] on email statistics predicts that an average user will send 35 emails and will receive 75 emails per day in 2012 (the study predicts the numbers to increase annually). In addition, membership in mailing lists[14] and Internet groups[15][16] is popular among Internet users, producing even more emails by normal email users. As an indication of the popularity of such services, Yahoo in 2010 announced[17] that 115 million unique users are collectively members of more than 10 million Yahoo Groups. Based on the mentioned statistics, we estimate that a conservative SWEET user can perform 35-70 web downloads per day, or make 10-20 interactive web connections, while ensuring unobservability of SWEET usage. Once again, as discussed in Section 5.1, we argue that unobservability is a concern only to special citizens; hence, we believe that normal citizens would use SWEET ignoring the possibilities of traffic analysis by a censor.

## 8. CONCLUSIONS

In this paper, we presented SWEET, a deployable system for unobservable communication with Internet destinations. SWEET works by tunneling network traffic through widely-used public email services such as Gmail, Yahoo Mail, and Hotmail. Unlike recently-proposed schemes that require a collection of ISPs to instrument router-level modifications in support of covert communications, our approach can be deployed through a small applet running at the user's end host, and a remote email-based proxy, simplifying deployment. Through an implementation and evaluation in a wide-area deployment, we find that while SWEET incurs some additional latency in communications, these overheads are low enough to be used for interactive accesses to web services. We feel our work may serve to accelerate deployment of censorship-resistant services in the wide area, guaranteeing high availability.

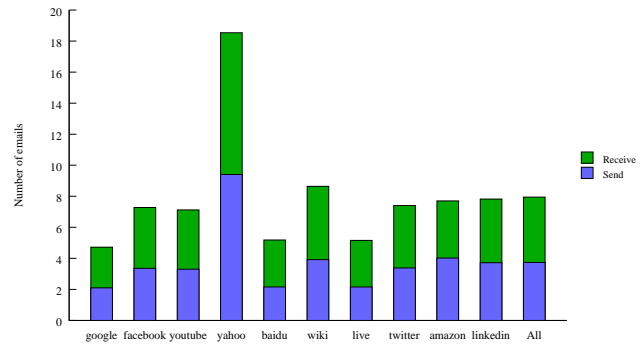

Figure 11: The number of emails sent and received by a SWEET client to get one of the websites from Alexa's top ten ranking.

---

[13] http://royal.pingdom.com/2011/01/12/internet-2010-in-numbers/
[14] http://gcc.gnu.org/lists.html
[15] http://groups.yahoo.com
[16] http://groups.google.com
[17] http://www.eweek.com/c/a/Search-Engines/Yahoo-Refreshes-Upgrades-Some-Products-675120/